\documentclass{Interspeech2024}

\usepackage{color}
\usepackage{multirow}
\usepackage{algorithm,algorithmic}
\usepackage{url}
\usepackage{amssymb,amsmath,bm}
\usepackage{textcomp}
\usepackage{cite}
\usepackage{CJKutf8}
\usepackage{hhline}
\usepackage{cite}
\usepackage{bbding}
\usepackage{pifont}
\usepackage{wasysym}
\usepackage{comment}
\usepackage{arydshln}

\def\L{{\cal L}}

\newcommand{\maru}[1]{\raise0.2ex\hbox{\textcircled{\scriptsize{#1}}}}

\interspeechcameraready

\title{Boosting Hybrid Autoregressive Transducer-based ASR with \\ Internal Acoustic Model Training and Dual Blank Thresholding}
\name[]{Takafumi}{Moriya}
\name[]{Takanori}{Ashihara}
\name[]{Masato}{Mimura}
\name[]{Hiroshi}{Sato}
\name[]{\\ Kohei}{Matsuura}
\name[]{Ryo}{Masumura}
\name[]{Taichi}{Asami}

\address{
  NTT Corporation, Japan
  }
\email{takafumi.moriya@ntt.com}


\begin{document}

\maketitle

\begin{abstract}
A hybrid autoregressive transducer (HAT) is a variant of neural transducer that models blank and non-blank posterior distributions separately. In this paper, we propose a novel internal acoustic model (IAM) training strategy to enhance HAT-based speech recognition. IAM consists of encoder and joint networks, which are fully shared and jointly trained with HAT. This joint training not only enhances the HAT training efficiency but also encourages IAM and HAT to emit blanks synchronously which skips the more expensive non-blank computation, resulting in more effective blank thresholding for faster decoding. Experiments demonstrate that the relative error reductions of the HAT with IAM compared to the vanilla HAT are statistically significant. Moreover, we introduce dual blank thresholding, which combines both HAT- and IAM-blank thresholding and a compatible decoding algorithm. This results in a 42-75\% decoding speed-up with no major performance degradation.
\end{abstract}

\section{Introduction}
\label{sec:intro}
%\vspace{-0.15cm}
End-to-end automatic speech recognition (E2E-ASR) has drawing attention in the ASR research community~\cite{jinyu2022survey,rohit2024suvey}. 
Several E2E-ASR models have been proposed,
e.g., connectionist temporal classification (CTC)~\cite{Graves2006},
attentional encoder-decoder (AED)~\cite{chorowski2014,vaswani2017att} and recurrent neural network-transducer (RNNT)~\cite{Graves2012}.
Recently, factorized E2E-ASR models have been introduced~\cite{Ehsan2020hat,lu2021MWER_HAT,Meng2022ModularHA,Meng2023jeit,xie2022factorized_nt,rui2023factorized_nt_adapt,Wu2024tSOT_FNT,xun2023factorized_aed,le2023BlankThresholding,hou2023CTCBlankThresholding}.
Given that RNNT is a promising technology for streaming ASR applications~\cite{Sainath2021AnES}, we focus on enhancing the factorized variant of RNNT.

Vanilla RNNT~\cite{Graves2012} is composed of encoder, prediction, and joint networks. The joint network uses a single distribution to model blank and non-blank probabilities. In inferencing, the joint network predicts a token from the distribution, and softmax computation is required at every frame, resulting in slower decoding as the vocabulary size increases. This poses a critical barrier to quick response, particularly for streaming ASR. 

To address this problem, we utilize the hybrid autoregressive transducer (HAT)~\cite{Ehsan2020hat} architecture, which factorizes the RNNT joint network into two heads that separately model the blank and non-blank probability distribution. While the factorization of RNNT typically aims to enhance language model (LM) integration~\cite{Ehsan2020hat,lu2021MWER_HAT,Meng2022ModularHA,Meng2023jeit,xie2022factorized_nt,rui2023factorized_nt_adapt,Wu2024tSOT_FNT}, we propose to use it for facilitating efficient decoding, as in~\cite{le2023BlankThresholding}. HAT first obtains the blank posterior and then decides whether to compute the non-blank probabilities at each frame, which is computationally expensive due to the large vocabulary size, assuming a manually preset threshold. This improves the decoding speed while retaining ASR performance. Moreover, in~\cite{hou2023CTCBlankThresholding}, this factorization was also applied to develop factorized CTC (FCTC). In this paper, we propose novel training and decoding approaches that benefit from the complementary nature of HAT and FCTC.

Several studies have reported that joint training with the CTC objective enhances ASR performance for AED and RNNT models~\cite{watanabe2017hybrid,boyer2021rnntctc,moritz2022merl_rnnt,moritz2023meta_rnnt}, by encouraging monotonic alignments between input speech and target labels. In this paper, we investigate whether joint training of HAT and CTC also yields synergistic benefits. We employ three types of CTC for joint training: 1) vanilla CTC~\cite{Graves2006} with a single distribution, 2) FCTC~\cite{hou2023CTCBlankThresholding} with separate blank and non-blank posterior distributions to match those of HAT, and 3) our proposed \textit{internal acoustic model} (IAM). IAM comprises encoder and joint networks, which are fully tied and jointly trained with HAT. 

By implementing IAM for joint training with HAT, we can take advantage of full parameter sharing when decoding. 
IAM also independently predicts blank and non-blank probabilities, similar to HAT, thereby enabling blank thresholding~\cite{hou2023CTCBlankThresholding}. 
In this paper, we introduce \textit{dual blank thresholding}, which combines HAT- and IAM-blank thresholding methods~\cite{le2023BlankThresholding,hou2023CTCBlankThresholding}; the goal is to skip non-blank posterior computation and so further enhance the decoding speed. Moreover, we explore a compatible decoding algorithm for dual blank thresholding that aims to mitigate the performance degradation caused by erroneous frame-skipping in blank thresholding.

Experiments demonstrate that all CTC objectives enhance HAT performance, with statistically significant relative error reductions compared to the vanilla HAT in both offline and streaming modes. Additionally, setting IAM within HAT enables the synchronous emission of blank symbols, unlike FCTC, which is separately optimized from the HAT decoder. Consequently, deploying our proposed HAT with IAM and dual blank thresholding, along with its compatible decoding algorithm, yields a 42-75\% increase in decoding speed without any significant degradation in ASR performance.

\begin{figure}[t]
  \begin{center}
    %\vspace{-0.3cm}
    \includegraphics[clip, width=6.5cm]{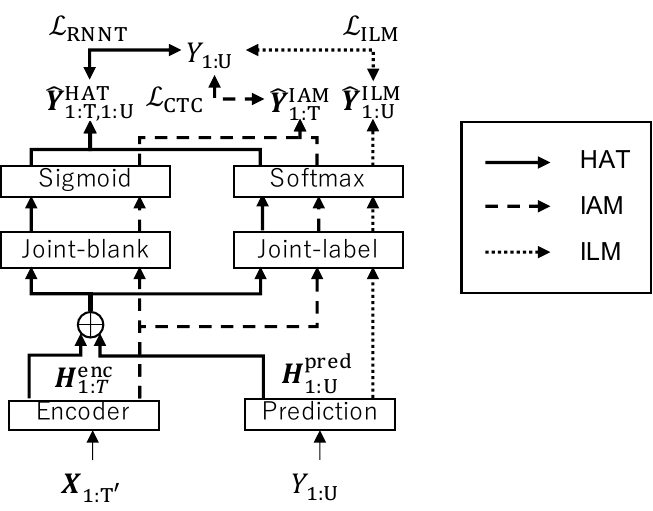}
    \vspace{-0.2cm}
    \caption{The architecture of the proposal: HAT with IAM and ILM. Solid, dashed, and dotted arrows show HAT, IAM, and ILM paths, respectively. If we zero out the prediction or encoder network output, it becomes IAM or ILM, respectively.}
  \label{fig:transducer}
  \end{center}
  \vspace{-0.8cm}
\end{figure}

\vspace{-0.025cm}
\section{ASR models}
\label{sec:modeling}
\vspace{-0.025cm}
Let $\bm{X}_{1:T^{\prime}} = \left[ \bm{x}_1, ..., \bm{x}_{T^{\prime}} \right]$ be the acoustic feature sequence with length-$T^{\prime}$, and $Y_{1:U} = \left[ y_1, ..., y_U \right]$ be a non-blank token sequence with length-$U$, where $y_u \in \{1, ..., K\}$. $K$ represents the number of tokens, including blank and non-blank symbols.

\vspace{-0.2cm}
\subsection{Neural transducer models}
\label{ssec:transducer}
\vspace{-0.1cm}
\subsubsection{RNNT}
\label{sssec:rnnt}
\vspace{-0.05cm}
RNNT~\cite{Graves2012} learns the mapping between sequences of different lengths. 
$\bm{X}_{1:T^{\prime}}$ is encoded and subsampled into $\bm{H}_{1:T}^{\text{enc}} = \left[ \bm{h}^{\text{enc}}_{1}, ..., \bm{h}^{\text{enc}}_{T} \right]$
of length-$T$ via encoder network $f^{\text{enc}}(\cdot)$.
$Y_{1:U}$ is also transformed into
$\bm{H}_{1:U}^{\text{pred}} = \left[ \bm{h}^{\text{pred}}_{1}, ..., \bm{h}^{\text{pred}}_{U} \right]$
via prediction network $f^{\text{pred}}(\cdot)$.
These encoded features are then fed to joint network $f^{\text{joint}}(\cdot)$ to obtain $\hat{\bm{y}}_{t,u} \in \mathbb{R}^{K}$, which contains both blank and non-blank posteriors within a single distribution. The above operations are defined as follows:
%\vspace{-0.2cm}
\begin{eqnarray}
\bm{h}^{\text{enc}}_{t} &=& f^{\text{enc}} (\bm{x}_{t^{\prime}}; \theta^{\text{enc}}), \\
\bm{h}^{\text{pred}}_{u} &=& f^{\text{pred}} (y_{u-1}; \theta^{\text{pred}}), \\
\hat{\bm{y}}_{t,u} &=& \text{Softmax} \left(f^{\text{joint}} (\bm{h}^{\text{enc}}_{t}, \bm{h}^{\text{pred}}_{u}; \theta^{\text{joint}}) \right),
%\end{aligned}\right.
\label{eq:rnnt}
%\vspace{-0.2cm}
\end{eqnarray}
where $\text{Softmax}(\cdot)$ means a softmax operation. 
RNNT outputs three dimensional tensor $\hat{\bm{Y}}^{\text{RNNT}}_{1:T,1:U} \in \mathbb{R}^{T \times U \times K}$ during training.
The learnable parameters $\theta^{\text{RNNT}} \triangleq [\theta^{\text{enc}}, \theta^{\text{pred}}, \theta^{\text{joint}}]$
are optimized using RNNT loss $\L_{\text{RNNT}}$~\cite{Graves2012}.

\vspace{-0.1cm}
\subsubsection{HAT}
\label{sssec:hat}
\vspace{-0.05cm}
Figure~\ref{fig:transducer} illustrates the HAT~\cite{Ehsan2020hat} architecture. 
HAT is a variant of RNNT and has two-head joint networks, which separately model blank probability $\hat{y}_{t,u}^{\text{blank}} \in \mathbb{R}^{1}$ and non-blank (label) probabilities $\hat{\bm{y}}_{t,u}^{\text{label}} \in \mathbb{R}^{K-1}$ following their different distributions. Thus, the joint network in Eq. (\ref{eq:rnnt}) is factorized into $f^{\text{joint-blank}}(\cdot)$ and $f^{\text{joint-label}}(\cdot)$. The joint network output is then concatenated. The above operations are defined as follows:
\begin{eqnarray}
\hat{y}_{t,u}^{\text{blank}} &=& \text{Sigmoid} \left(f^{\text{joint-blank}} (\bm{h}^{\text{enc}}_{t}, \bm{h}^{\text{pred}}_{u}; \theta^{\text{joint-blank}}) \right), \label{eq:hat_blank}\\
\hat{\bm{y}}_{t,u}^{\text{label}} &=& \text{Softmax} \left(f^{\text{joint-label}} (\bm{h}^{\text{enc}}_{t}, \bm{h}^{\text{pred}}_{u}; \theta^{\text{joint-label}}) \right), \label{eq:hat_label}\\
\hat{\bm{y}}_{t,u} &=& [ \ \hat{y}_{t,u}^{\text{blank}} ; \ (1 - \hat{y}_{t,u}^{\text{blank}}) \cdot \hat{\bm{y}}_{t,u}^{\text{label}} \ ],
\label{eq:hat}
\end{eqnarray}
where $\text{Sigmoid}(\cdot)$ means a sigmoid function. HAT outputs $\hat{\bm{Y}}^{\text{HAT}}_{1:T,1:U} \in \mathbb{R}^{T \times U \times K}$ in the same format as that of RNNT.
The learnable parameters $\theta^{\text{HAT}} \triangleq [\theta^{\text{enc}}, \theta^{\text{pred}}, \theta^{\text{joint-blank}}, \theta^{\text{joint-label}}]$
are optimized using $\L_{\text{RNNT}}$.

% CTC
\vspace{-0.1cm}
\subsection{CTC models}
\label{ssec:ctc}
\vspace{-0.05cm}
\subsubsection{Vanilla CTC}
\label{sssec:ctc}
\vspace{-0.05cm}
CTC~\cite{Graves2006} also learns the alignments between sequences of different lengths. 
In this work, an additional linear layer $f^{\text{linear}}(\cdot)$ is stacked on top of the shared encoder network with each neural transducer. 
CTC predictions $\hat{\bm{y}}_{t} \in \mathbb{R}^{K}$ are obtained as follows:
\begin{eqnarray}
\hat{\bm{y}}_{t} &=& \text{Softmax} \left(f^{\text{linear}} (\bm{h}^{\text{enc}}_{t}; \theta^{\text{linear}}) \right).
\label{eq:ctc}
\end{eqnarray}
CTC outputs two dimensional matrix $\hat{\bm{Y}}^{\text{CTC}}_{1:T} \in \mathbb{R}^{T \times K}$ during training.
The learnable parameters $\theta^{\text{CTC}} \triangleq [\theta^{\text{enc}}, \theta^{\text{linear}}]$ is optimized using CTC loss $\L_{\text{CTC}}$~\cite{Graves2006}.

\subsubsection{Factorized CTC (FCTC)}
\label{sssec:fctc}
\vspace{-0.05cm}
FCTC was recently proposed in~\cite{hou2023CTCBlankThresholding}. It also separately models blank probability $\hat{y}_{t}^{\text{blank}} \in \mathbb{R}^{1}$ and non-blank probabilities $\hat{\bm{y}}_{t}^{\text{label}} \in \mathbb{R}^{K-1}$, similar to HAT. Thus, the linear layer in Eq. (\ref{eq:ctc}) can be factorized into $f^{\text{linear-blank}}(\cdot)$ and $f^{\text{linear-label}}(\cdot)$, and both linear layers are stacked on top of the shared encoder network. The output of the linear layers is concatenated as follows:
\begin{eqnarray}
\hat{y}_{t}^{\text{blank}} &=& \text{Sigmoid} \left(f^{\text{linear-blank}} (\bm{h}^{\text{enc}}_{t}; \theta^{\text{linear-blank}}) \right), \\
\hat{\bm{y}}_{t}^{\text{label}} &=& \text{Softmax} \left(f^{\text{linear-label}} (\bm{h}^{\text{enc}}_{t}; \theta^{\text{linear-label}}) \right), \\
\hat{\bm{y}}_{t} &=& [ \ \hat{y}_{t}^{\text{blank}} ; \ (1 - \hat{y}_{t}^{\text{blank}}) \cdot \hat{\bm{y}}_{t}^{\text{label}} \ ].
\label{eq:fctc}
\end{eqnarray}
FCTC outputs $\hat{\bm{Y}}^{\text{FCTC}}_{1:T} \in \mathbb{R}^{T \times K}$ in the same shape as the vanilla CTC. 
The learnable parameters $\theta^{\text{FCTC}} \triangleq [\theta^{\text{enc}}, \theta^{\text{linear-blank}}, \theta^{\text{linear-label}}]$
are optimized using $\L_{\text{CTC}}$.

\vspace{-0.1cm}
\subsubsection{Proposed internal acoustic model (IAM) within HAT}
\label{sssec:iam}
\vspace{-0.05cm}
Vanilla CTC and FCTC require additional parameters for each linear output layer. To create simple ASR systems, we introduce IAM within HAT, see Fig.~\ref{fig:transducer}. 
Implementing IAM is very straightforward. We consistently replace the prediction network output $\bm{h}^{\text{pred}}_{u}$ with zero vector $\bm{0}$ in Eq. (\ref{eq:hat_blank}) and (\ref{eq:hat_label}) for IAM\footnote{The IAM framework can also be applied to RNNT, and we evaluate the vanilla CTC-like IAM within RNNT in our experiments.}. 
Thus, IAM acts as the counterpart to the internal LM (ILM) factorization~\cite{Ehsan2020hat}, which conversely replaces $\bm{h}^{\text{enc}}_{t}$ with $\bm{0}$, also see Fig.~\ref{fig:transducer}. 
The number of IAM output distributions equals that of HAT.
IAM within HAT also outputs two-dimensional matrix $\hat{\bm{Y}}^{\text{IAM}}_{1:T} \in \mathbb{R}^{T \times K}$ like FCTC. 
The learnable parameters $\theta^{\text{IAM}} \triangleq [\theta^{\text{enc}}, \theta^{\text{joint-blank}}, \theta^{\text{joint-label}}]$ are optimized using $\L_{\text{CTC}}$.

\vspace{-0.1cm}
%\subsection{Training and decoding}
\subsection{Training}
\vspace{-0.05cm}
\label{trainingdecoding}
We perform joint training, which combines $\L_{\text{RNNT}}$ with $\L_{\text{CTC}}$ and ILM training loss $\L_{\text{ILM}}$~\cite{Ehsan2020hat,meng2021ilmt} as follows:
\begin{equation}
\L_{\text{Joint}} = \L_{\text{RNNT}} + \alpha \L_{\text{CTC}} + \beta \L_{\text{ILM}},
\label{eq:trainloss}
\end{equation}
where $\alpha$ and $\beta$ are the weights of $\L_{\text{CTC}}$ and $\L_{\text{ILM}}$, respectively.

\vspace{-0.1cm}
\subsection{Decoding with blank thresholding}
\label{ssec:threshold}
\vspace{-0.05cm}
Neural transducer and CTC models emit blank symbols more often than non-blank symbols. Hence, calculating non-blank probabilities can be dropped for the majority of time frames as we discuss below, while RNNT and vanilla CTC must perform this computation at every frame. 
In this paper, we investigate blank thresholding techniques~\cite{le2023BlankThresholding,tian2021fsr,Wang2022AcceleratingRT,yang2023blank_regularized_ctc,hou2023CTCBlankThresholding} to achieve efficient decoding. 

\vspace{-0.1cm}
\subsubsection{HAT-blank thresholding}
\label{sssec:hat_threshold}
\vspace{-0.05cm}
Figure~\ref{fig:threshold} (a) illustrates HAT-blank thresholding~\cite{le2023BlankThresholding}. HAT predicts $\hat{y}_{t}^{\text{blank}}$ conditioned on, not only $\bm{h}^{\text{enc}}_{t}$, but also $\bm{h}^{\text{pred}}_{u}$, which enforces autoregressive decoding. 
If $\hat{y}_{t}^{\text{blank}} < \text{Sigmoid} (\lambda^{\text{HAT}})$, we can skip the computation of non-blank probability $\hat{\bm{y}}_{t}^{\text{label}}$, where $\lambda^{\text{HAT}}$ is a manually defined threshold. HAT-blank thresholding reduces the computation time required for $\hat{\bm{y}}_{t}^{\text{label}}$ in the HAT decoder, but it requires slow frame-by-frame operation due to the autoregressive nature. 

\vspace{-0.1cm}
\subsubsection{CTC-blank thresholding}
\label{sssec:ctc_threshold}
\begin{figure}[t]
  \begin{center}
    %\vspace{-0.3cm}
    \includegraphics[clip, width=7.9cm]{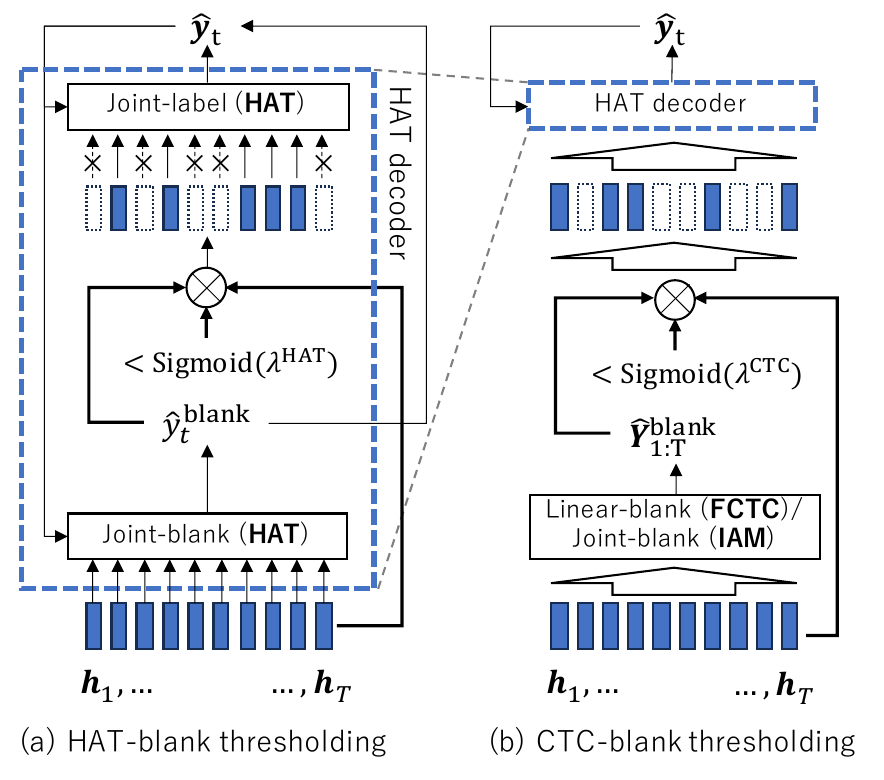}
    \vspace{-0.3cm}
    \caption{Schematic diagram of (a) autoregressive HAT- and (b) non-autoregressive CTC-blank thresholding approaches. $\lambda^{*}$ is the threshold hyperparameter.}
  \label{fig:threshold}
  \end{center}
  \vspace{-0.9cm}
\end{figure}

\vspace{-0.05cm}
The CTC-blank thresholding~\cite{hou2023CTCBlankThresholding,tian2021fsr,Wang2022AcceleratingRT,yang2023blank_regularized_ctc}, depicted in Fig.~\ref{fig:threshold} (b), can also skip the computation of non-blank posterior $\hat{\bm{y}}_{t}^{\text{label}}$ in HAT if $\hat{y}_{t}^{\text{blank}} < \text{Sigmoid} (\lambda^{\text{CTC}})$. In this paper, $\hat{\bm{Y}}_{1:T}^{\text{blank}}$ is obtained from FCTC or IAM of HAT for faster thresholding. The difference from HAT-blank thresholding is that CTC-blank thresholding can perform thresholding for all frames in parallel.
This is because CTC, which performs non-autoregressive decoding unlike neural transducers, can predict $\hat{\bm{Y}}_{1:T}^{\text{blank}}$ for all frames in parallel. This enables discarding $\bm{h}_{t}^{\text{enc}}$ for most $t$ before inputting them into the transducer decoder.
%The CTC-blank thresholding can also be applied to streaming ASR with a chunkwise encoder~\cite{chen2021lcconformer,yangyang2022emformer}.

\vspace{-0.1cm}
\subsubsection{Proposed dual blank thresholding}
\label{sssec:dual_threshold}
\vspace{-0.1cm}
In~\cite{le2023BlankThresholding,hou2023CTCBlankThresholding}, the authors applied each blank thresholding method individually. In this paper, we introduce dual blank thresholding which combines these two methods. Our proposed thresholding employs CTC-blank thresholding to eliminate unnecessary encoder outputs. Then, it passes the remaining encoder outputs to the HAT-decoder, where slow but more reliable HAT-blank thresholding is applied. We expect that the proposed two-step blank thresholding will further enhance the decoding speed.

\vspace{-0.1cm}
\subsubsection{Compatible decoding algorithm for frame-skipping}
\label{sssec:decoding}
\vspace{-0.1cm}
The discard of encoder outputs by HAT- or CTC-blank thresholding may degrade the ASR performance. We investigate two popular decoding algorithms, which are alignment-length synchronous decoding (ALSD) and time-synchronous decoding (TSD)~\cite{saon2020alsd}, each with blank thresholding. ALSD explores the top hypotheses, each having higher scores along the shortest path.%\footnote{For streaming model, we perform chunkwise deoding, and thus ALSD operates also as chunkwise processing.}. 
TSD finds the best hypotheses in the search space by performing label expansion for each frame. 
We investigate these decoding algorithms to mitigate the performance degradation caused by erroneous blank thresholding.

\vspace{-0.1cm}
\section{Experiments}

\vspace{-0.1cm}
\subsection{Data}
\label{ssec:data}
\vspace{-0.1cm}
We evaluated our proposed approach on TED-LIUM release-2 (TLv2)~\cite{tedlium2} and LibriSpeech~\cite{librispeech}.
The datasets contain speech samples totaling 210 and 960 hours, respectively.
In this paper, we adopted Byte Pair Encoding~\cite{sennrich2016bpe} for tokenization and used 2000 subwords for TLv2 and 5000 subwords for LibriSpeech, respectively.
Data augmentation methods~\cite{povey2015sp,specaugment} were applied to both datasets during training. 
The datasets were preprocessed following the corresponding Kaldi and ESPnet recipes~\cite{povey2011kaldi,espnet}.

\begin{table}[t]
  \caption{Comparisons of WER [\%] on TLv2 test set using \textbf{offline} models with different CTC types and $\alpha$ in Eq. (\ref{eq:trainloss}).}
  \vspace{-3mm}
  \label{tab:summary_offline}
  \centering
  \scalebox{0.88}[0.88]{
\begin{tabular}{c|l|l|c|c@{ }@{ }c@{ }@{ }c@{ }@{ }c}
\hline
\multirow{2}{*}{ID} & \multirow{2}{*}{Model} & \multirow{2}{*}{CTC type} & \multicolumn{5}{c}{CTC loss weight $\alpha$}                                           \\
    &                        &                        & 0.00                  & 0.25          & 0.50 & 0.75          & 1.00 \\ \hline
OC1 & CTC                    & -                      & -                  & -             & -    & -             & 7.90    \\ \hline
OC2 & FCTC                   & -                      & -                  & -             & -    & -             & 7.76    \\ \hline
OR1 & \multirow{3}{*}{RNNT}  & CTC                    & \multirow{3}{*}{7.56} & \textbf{6.89} & 7.00 & 7.06          & 7.12 \\
OR2 &                        & FCTC                   &                       & \textbf{6.96} & 7.03 & 7.09          & 7.14 \\
OR3 &                        & IAM (CTC)                &                       & \textbf{6.95} & 7.01 & 7.14          & 7.22 \\ \hline
OH1 & \multirow{3}{*}{HAT}   & CTC                      & \multirow{3}{*}{7.65} & 7.41          & 7.20 & \textbf{6.98} & 7.23 \\
OH2 &                        & FCTC                     &                       & 7.36          & 7.19 & \textbf{7.08} & 7.13 \\
OH3 &                        & IAM (FCTC)               &                       & 7.30          & 7.19 & \textbf{6.92} & 7.09 \\ \hline
\end{tabular}
}
\vspace{-0.1cm}
\end{table}

\begin{table}[t]
  \caption{Comparisons of WER [\%] on TLv2 test set using \textbf{streaming} models with different CTC types and $\alpha$ in Eq. (\ref{eq:trainloss}).}
  \vspace{-3mm}
  \label{tab:summary_streaming}
  \centering
  \scalebox{0.88}[0.88]{
\begin{tabular}{c|l|l|c|c@{ }@{ }c@{ }@{ }c@{ }@{ }c}
\hline
\multirow{2}{*}{ID} & \multirow{2}{*}{Model} & \multirow{2}{*}{CTC type} & \multicolumn{5}{c}{CTC loss weight $\alpha$}                                           \\
     &                       &                           & 0.00                  & 0.25          & 0.50 & 0.75          & 1.00 \\ \hline
SC1  & CTC                    & -                         & -                 & -             & -    & -             & 10.68    \\ \hline
SC2  & FCTC                   & -                         & -                 & -             & -    & -             & 10.22    \\ \hline
SR1  &\multirow{3}{*}{RNNT}  & CTC                      & \multirow{3}{*}{9.41} & \textbf{8.46}  & 8.53 & 8.70          & 8.78 \\
SR2  &                       & FCTC                     &                       & \textbf{8.41}  & 8.66 & 8.75        & 8.81  \\
SR3  &                       & IAM (CTC)                &                       & \textbf{8.67}  & 8.80 & 9.14          & 9.30 \\ \hline
SH1  &\multirow{3}{*}{HAT}   & CTC                      & \multirow{3}{*}{9.36} & 8.85           & \textbf{8.60} & 8.76          & 8.84 \\
SH2  &                       & FCTC                     &                       & 8.93           & 8.71 & \textbf{8.66} & 8.76 \\
SH3  &                       & IAM (FCTC)               &                       & 8.96  & \textbf{8.44} & 8.55          & 8.79 \\ \hline
\end{tabular}
}
\vspace{-0.5cm}
\end{table}

\vspace{-0.1cm}
\subsection{System configuration}
\label{ssec:system}
\vspace{-0.1cm}

The input feature was an 80-dimensional log Mel-filterbank extracted every 10 ms. 
We adopted a Conformer-transducer (L)~\cite{anmol2020conformer} with a kernel size of 15. 
For the streaming system, we replaced depthwise convolution and batch normalization with causal depthwise convolution and layer normalization, respectively.
We utilized two-layer 2-dimensional convolution layers followed by 17 Conformer blocks, where both stride sizes in the max-pooling layers were set to $2 \times 2$, resulting in a 30ms look-ahead requirement.
The prediction network had a 640-dimensional long short-term memory layer followed by the joint network. 
For the vanilla CTC or FCTC branch, the decoder was additionally stacked on top of the shared encoder with neural transducer models.
Streaming models were trained using an attention mask,
and chunkwise decoding was performed as in~\cite{chen2021lcconformer}.
Both the history and current chunk sizes were set to 600ms, so the algorithmic latency was 630ms. 

All models were trained from scratch using Eq. (\ref{eq:trainloss}) with the Noam learning rate scheduler, AdamW optimizer, and 25k warm-up steps~\cite{vaswani2017att,loshchilov2019adamw} for a total of 100 epochs for TLv2 and 50 epochs for LibriSpeech.
We investigated $\alpha = \{0.00, 0.25, 0.50, 0.75, 1.00\}$; $\beta$ was set to 0.1. 
For decoding, executed on an Intel Xeon Gold 6338 2.00GHz CPU, we adopted ALSD or TSD~\cite{saon2020alsd} with a beam size of 8 for neural transducers and greedy search for CTCs.
We evaluated performance in terms of word error rate (WER).
We also measured decoding efficiency in terms of non-blank percentage (NBP: \textit{non-blank frame length / encoder output length}) for CTC-blank thresholding, joint network call ratio (JCR: \textit{non-blank call times / blank call times}) for HAT-blank thresholding, and real-time factor (RTF: \textit{decoding time / data time}) for decoding speed.

\vspace{-0.15cm}
\subsection{Results}
\label{ssec:result}
\vspace{-0.1cm}

\subsubsection{Effectiveness of CTC objectives for HAT training}
\label{sssec:result_ctc}
\vspace{-0.1cm}

First, we investigated the effectiveness of incorporating CTC objectives for HAT training in TLv2. Table~\ref{tab:summary_offline} presents the WERs of offline ASR models. 
``CTC type'' indicates CTC architecture. 
We can see that the WERs of standalone neural transducers ($\alpha = 0.00$) are superior to those of standalone CTCs. Applying CTC objectives to each neural transducer training ($\alpha > 0$) resulted in improvements in both RNNT and HAT. 
RNNT with $\alpha = 0.25$ and HAT with $\alpha = 0.75$ achieved the best WERs. We performed the MAPSSWE significance test~\cite{mapsswe}, and the differences from $\alpha = 0.00$ were statistically significant, $p < 0.001$. 

Table~\ref{tab:summary_streaming} shows the WERs of streaming ASR models.
RNNT with $\alpha = 0.25$ and HAT with larger $\alpha$, i.e., 0.50-0.75, achieved the best WERs, and the differences from $\alpha = 0.00$ were also statistically significant ($p < 0.001$). 
The above results indicate that smaller $\alpha$ is suitable for RNNT, while larger $\alpha$ is more suitable for HAT. Since HAT trains blank and non-blank distributions separately, the alignment information provided from CTC may be more helpful for HAT training to obtain accurate alignment than for RNNT with a single distribution.

\begin{figure}[t]
  \begin{center}
    %\vspace{-0.3cm}
    \includegraphics[clip, width=8.0cm]{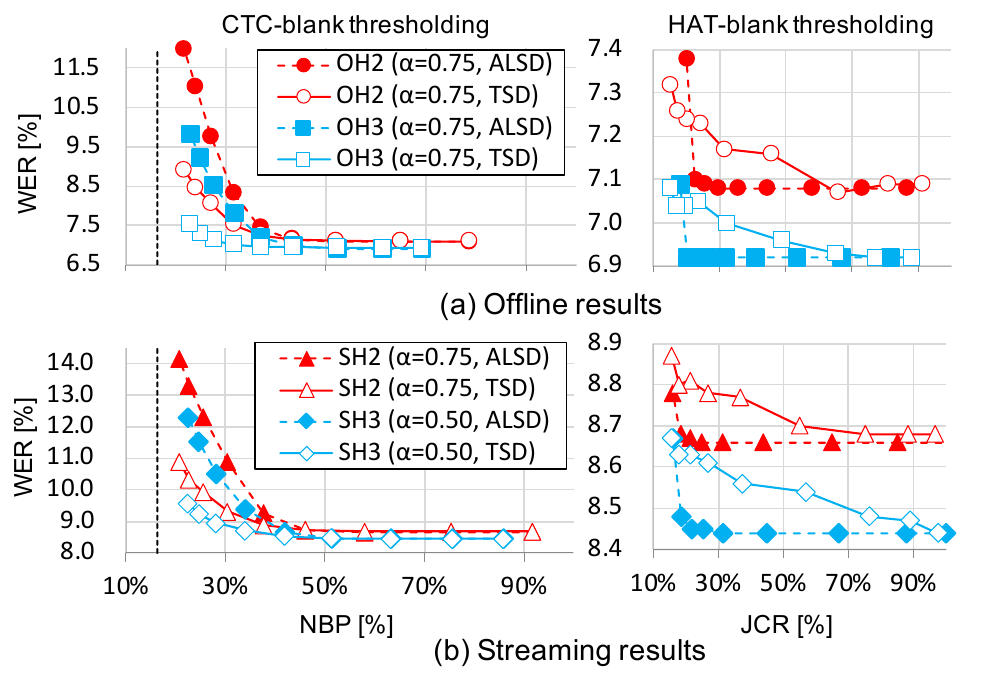}
    \vspace{-0.7cm}
    \caption{WER versus NBP/JCR curves. The lower-left region represents better thresholding and decoding algorithms. System IDs in the legends correspond to those in Table~\ref{tab:summary_offline} and~\ref{tab:summary_streaming}. }
%    \caption{Aggregated WER vs. NBP/JCR curves. The lower-left region represents better thresholding and decoding algorithms. System IDs in the legends correspond to those in Table~\ref{tab:summary_offline} and~\ref{tab:summary_streaming}. }
  \label{fig:results_threshold}
  \end{center}
  \vspace{-0.55cm}
\end{figure}

\begin{table}[t]
  \caption{Summary of the results using best configurations on TLv2 test set. System IDs correspond to those in Table~\ref{tab:summary_offline} and~\ref{tab:summary_streaming}. $^{\dag}$In dual blank thresholding, the JCR is larger because HAT-blank thresholding is applied after CTC-blank thresholding.}
  \vspace{-3mm}
  \label{tab:summary_tedlium2}
  \centering
  \scalebox{0.95}[0.95]{
\begin{tabular}{l|c|c@{ }c@{ }c@{ }@{ }c@{ }c@{ }@{ }c}
\hline
\multirow{2}{*}{ID}               & \multirow{2}{*}{\begin{tabular}[c]{@{}c@{}}WER\\ {[}\%{]}\end{tabular}} & \multirow{2}{*}{\begin{tabular}[c]{@{}c@{}}Dec.\\ Alg.\end{tabular}} & \multirow{2}{*}{$\lambda^{\text{CTC}}$} & \multirow{2}{*}{$\lambda^{\text{HAT}}$} & \multirow{2}{*}{\begin{tabular}[c]{@{}c@{}}NBP\\ {[}\%{]}\end{tabular}} & \multirow{2}{*}{\begin{tabular}[c]{@{}c@{}}JCR\\ {[}\%{]}\end{tabular}} & \multirow{2}{*}{RTF} \\
                                  &                                                                         &                                                                      &                         &                         &                                                                         &                                                                         &                      \\ \hline
\multirow{5}{*}{\begin{tabular}[c]{@{}l@{}}OH3:\\ HAT+IAM\\ ($\alpha = 0.75$)\end{tabular}} & 6.92                                                                    & ALSD                                                                 & -                       & -                       & 100                                                                     & 100                                                                     & \underline{0.258}                 \\
                                  & 6.92                                                                    & ALSD                                                                 & -                       & 2                       & 100                                                                     & 20                                                                      & 0.097                 \\
                                  & 6.95                                                                    & TSD                                                                  & 8                       & -                       & 37                                                                      & 100                                                                     & 0.094                 \\
                                  & 6.92                                                                    & ALSD                                                                 & 12                      & 2                       & 52                                                                      & $\text{33}^{\dag}$                                                                     & \textbf{0.072}                \\
                                  & 6.94                                                                    & TSD                                                                  & 8                       & 14                      & 37                                                                      & $\text{98}^{\dag}$                                                                      & 0.088                 \\ \hline
\multirow{5}{*}{\begin{tabular}[c]{@{}l@{}}SH3:\\ HAT+IAM\\ ($\alpha = 0.50$)\end{tabular}} & 8.44                                                                    & ALSD                                                                 & -                       & -                       & 100                                                                     & 100                                                                     & \underline{0.365}                \\
                                  & 8.48                                                                    & ALSD                                                                 & -                       & 2                       & 100                                                                     & 22                                                                      & 0.243                 \\
                                  & 8.49                                                                    & TSD                                                                  & 8                      & -                       & 42                                                                      & 100                                                                     & 0.242                 \\
                                  & 8.48                                                                    & ALSD                                                                 & 10                      & 2                       & 52                                                                      & $\text{35}^{\dag}$                                                                      & \textbf{0.211}                 \\
                                  & 8.49                                                                    & TSD                                                                  & 8                      & 14                      & 42                                                                      & $\text{99}^{\dag}$                                                                      & 0.235                 \\ \hline
\end{tabular}
}
\vspace{-0.45cm}
\end{table}

\vspace{-0.225cm}
\subsubsection{Effectiveness of blank thresholding for faster decoding}
\label{sssec:result_blank_thresholding}
\vspace{-0.1cm}
Figure~\ref{fig:results_threshold} (a) and (b) illustrate the relationship between aggregated WER and NBP/JCR with each blank thresholding approach for offline and streaming models, respectively. We evaluated the combinations of CTC (left) and HAT (right) blank thresholding methods and decoding algorithms on TLv2 using the best systems in Table~\ref{tab:summary_offline} and~\ref{tab:summary_streaming}. Dashed and solid lines indicate the results achieved with ALSD and TSD, respectively. 
The black dashed line means the oracle NBP (=16\%), which is denoted as the number of ground-truth tokens divided by encoder output lengths.
The thresholds, i.e., $\lambda^{\text{HAT}}$ and $\lambda^{\text{CTC}}$, were set in increments of 2 from 0 to 16, plotted from left to right in each curve. We can see that as threshold $\lambda^{*}$ decreases, NBP and JCR for decoding efficiency improve, but WERs degrade. We found that TSD mitigates the degradation in WER caused by CTC-blank thresholding with smaller $\lambda^{\text{CTC}}$, while ALSD demonstrates robustness against HAT-blank thresholding with smaller $\lambda^{\text{HAT}}$. Interestingly, IAM in HAT exhibits less degradation than FCTC for CTC-blank thresholding in both offline and streaming modes. This is probably because IAM was jointly trained with HAT and shared all network parameters. Therefore, the blank emission timings of IAM can be synchronized with those of HAT more effectively than FCTC.

Table~\ref{tab:summary_tedlium2} presents the results from the systems with the best configurations. Here, we evaluated our proposed dual blank thresholding, which activates both $\lambda^{\text{HAT}}$ and $\lambda^{\text{CTC}}$. 
We can see that the proposed dual blank thresholding with ALSD achieves a decoding speed-up of 72\% for the offline system and 42\% for the streaming system compared to HAT without any blank thresholding (underlined). The smaller RTF improvements for the streaming models compared to the offline models are due to the relatively poor performance of streaming IAM blank prediction and the lower throughput associated with the for-loop used in chunkwise decoding.

\begin{table}[t]
  \caption{Comparisons of WER [\%] on LibriSpeech test sets using offline and streaming models with best configurations. RNNT and HAT were jointly trained with each IAM if $\alpha > 0$.}
  \vspace{-3mm}
  \label{tab:summary_librispeech}
  %\centering
  \hspace{-0.2cm} \scalebox{0.87}[0.87]{
\begin{tabular}{@{ }ll|c|c@{ }c|c@{ }c@{ }c@{ }c@{ }c@{ }c@{ }}
\hline
\multicolumn{2}{l|}{\multirow{2}{*}{Model}}       & \multirow{2}{*}{$\alpha$} & \multicolumn{2}{c|}{WER {[}\%{]}} & \multirow{2}{*}{\begin{tabular}[c]{@{}c@{}}Dec.\\ Alg.\end{tabular}} & \multirow{2}{*}{$\lambda^{\text{CTC}}$} & \multirow{2}{*}{$\lambda^{\text{HAT}}$} & \multirow{2}{*}{\begin{tabular}[c]{@{}c@{}}NBP\\ {[}\%{]}\end{tabular}} & \multirow{2}{*}{\begin{tabular}[c]{@{}c@{}}JCR\\ {[}\%{]}\end{tabular}} & \multirow{2}{*}{RTF} \\
\multicolumn{2}{l|}{}                             &                        & clean           & other           &                                                                      &                        &                        &                                                                         &                                                                         &                      \\ \hline
\multicolumn{1}{@{ }l@{ }|}{\multirow{7}{*}{\rotatebox{90}{Offline}}}  & CTC  & -                      & 3.21            & 7.99            & greedy                                                               & -                      & -                      & 100                                                                     & -                                                                       & 0.071                 \\
\multicolumn{1}{@{ }l@{ }|}{}                      & FCTC & -                      & 3.28            & 7.74            & greedy                                                               & -                      & -                      & 100                                                                     & -                                                                       & 0.072                 \\
\multicolumn{1}{@{ }l@{ }|}{}                      & RNNT & 0.25                   & 2.86            & 7.12            & ALSD                                                                 & -                      & -                      & 100                                                                     & 100                                                                     & 0.332                 \\ \cline{2-11} 
\multicolumn{1}{@{ }l@{ }|}{}                      & HAT  & 0.00                   & 3.06                & 7.55                & ALSD                                                                 & -                      & -                      & 100                                                                     & 100                                                                     & 0.342                 \\
\multicolumn{1}{@{ }l@{ }|}{}                      &      & 0.75                   & 2.94            & 7.21            & ALSD                                                                 & -                      & -                      & 100                                                                     & 100                                                                     & \underline{0.344}                 \\
\multicolumn{1}{@{ }l@{ }|}{}                      &      & 0.75                   & 2.98            & 7.28            & ALSD                                                                 & 14                     & 2                      & 46                                                                      & 28                                                                      & 0.098                 \\
\multicolumn{1}{@{ }l@{ }|}{}                      &      & 0.75                   &  2.98               & 7.26                & TSD                                                                  &  10                     & 0                     & 31                                                                        & 29                                                                        & \textbf{0.086}                 \\ \hline
\multicolumn{1}{@{ }l@{ }|}{\multirow{7}{*}{\rotatebox{90}{Streaming}}} & CTC  & -                      & 5.18            & 12.88           & greedy                                                               & -                      & -                      & 100                                                                     & -                                                                       & 0.230                 \\
\multicolumn{1}{@{ }l@{ }|}{}                      & FCTC & -                      & 4.86            & 12.20           & greedy                                                               & -                      & -                      & 100                                                                     & -                                                                       & 0.232                 \\
\multicolumn{1}{@{ }l@{ }|}{}                      & RNNT & 0.25                   & 4.03            & 10.57           & ALSD                                                                 & -                      & -                      & 100                                                                     & 100                                                                     & 0.427                 \\ \cline{2-11} 
\multicolumn{1}{@{ }l@{ }|}{}                      & HAT  & 0.00                   & 4.34            & 10.86           & ALSD                                                                 & -                      & -                      & 100                                                                     & 100                                                                     & 0.454                 \\
\multicolumn{1}{@{ }l@{ }|}{}                      &      & 0.50                   & 3.91            & 10.44           & ALSD                                                                 & -                      & -                      & 100                                                                     & 100                                                                     & \underline{0.448}                 \\
\multicolumn{1}{@{ }l@{ }|}{}         &      & 0.50                   & 3.91                & 10.42               & ALSD                                                                 & 12                       & 2                       & 53                                                                        & 29                                                                         & 0.258                 \\
\multicolumn{1}{@{ }l@{ }|}{}         &      & 0.50                   & 3.95                & 10.47                & TSD                                                                  & 12                      &  0                    &  53                                                                       & 20                                                                       & \textbf{0.246}                 \\ \hline
\end{tabular}
}
  \vspace{-0.55cm}
\end{table}

\vspace{-0.25cm}
\subsubsection{Validity on another dataset (LibriSpeech)}
\label{sssec:result_librispeech}
\vspace{-0.15cm}
Finally, we evaluated our proposed approaches on LibriSpeech, and the results are shown in Table~\ref{tab:summary_librispeech}. The oracle NBP was 14\%.
The joint training of HAT and our proposed IAM led to WER improvements. 
Our proposed dual blank thresholding with TSD achieved 75\% faster decoding for the offline system and 45\% for the streaming system compared to HAT without any blank thresholding (underlined). 
For the LibriSpeech task, HAT-blank thresholding using TSD with smaller $\lambda^{\text{HAT}}$ also worked. This is probably because TSD is suitable for short-length speech samples without pause duration.
Interestingly, the RTF of HAT, jointly trained with IAM and using dual blank thresholding with TSD, matched that of vanilla CTC while HAT performed beam search decoding. 

\vspace{-0.225cm}
\section{Conclusion}
\vspace{-0.1cm}
We explored two approaches to enhance HAT-based ASR: 1) joint training with various CTC objectives improved the performance of HAT-based ASR with statistical significance, and 2) dual blank thresholding using HAT and IAM led to 42-75\% faster decoding. With regard to inferencing, the IAM demonstrated that fully sharing the parameters with HAT brought the blank emission timings closer together resulting competitive decoding speeds with CTC greedy search. Furthermore, we found an appropriate decoding algorithm that mitigated the performance degradation caused by erroneous blank thresholding. 
\clearpage

\bibliographystyle{IEEEtran}
\bibliography{mybib}

\end{document}